\documentstyle[osa]{revtex}
\begin{document}
\title{Exact Solution of the Energy Shift in each Quantum Mechanical Energy Levels
in a One Dimensional Symmetrical Linear Harmonic Oscillator \thinspace }
\author{Hendry Izaac Elim\thanks{%
email: hendry202@cyberlib.itb.ac.id}$^{)1)}$}
\author{1)\thinspace \thinspace Department of Physics,\thinspace Pattimura
University, Ambon,\thinspace \thinspace Indonesia\thinspace \thinspace
\thinspace \thinspace}
\date{January 7, 1999}
\maketitle

\begin{abstract}
\thinspace \thinspace \thinspace \thinspace \thinspace \thinspace \thinspace
\thinspace \thinspace \thinspace \thinspace \thinspace \thinspace \thinspace
\thinspace \thinspace \thinspace \thinspace \thinspace \thinspace \thinspace
\thinspace \thinspace \thinspace \thinspace \thinspace \thinspace \thinspace
\thinspace \thinspace \thinspace \thinspace \thinspace \thinspace \thinspace
\thinspace \thinspace \thinspace \thinspace \thinspace \thinspace \thinspace
\thinspace \thinspace \thinspace \thinspace \thinspace \thinspace \thinspace
\thinspace \thinspace \thinspace \thinspace \thinspace \thinspace \thinspace
\thinspace \thinspace \thinspace \thinspace \thinspace \thinspace \thinspace
\thinspace \thinspace \thinspace \thinspace \thinspace \thinspace \thinspace
\thinspace \thinspace \thinspace \thinspace \thinspace \thinspace \thinspace
\thinspace \thinspace \thinspace \thinspace \thinspace \thinspace \thinspace
\thinspace \thinspace \thinspace \thinspace \thinspace \thinspace \thinspace
\thinspace \thinspace \thinspace \thinspace \thinspace \thinspace \thinspace
\thinspace \thinspace \thinspace \thinspace \thinspace \thinspace \thinspace
\thinspace {\bf Abstract}

An exact solution of the energy shift in each quantum mechanical energy
levels in a one dimensional symmetrical linear harmonic oscillator has been
investigated. The solution we have used here is firstly derived by
manipulating Schr\"{o}dinger differential equation to be confluent
hypergeometric differential equation. The final exact numerical results of
the energy shifts are then found by calculating the final analytical
solution of the confluent hypergeometric equation with the use of a software
(Mathcad Plus 6.0) or a program programmed by using Turbo Pascal 7.0. We
find that the results of the energy shift in our exact solution method are
almost the same as that in Barton et. al. approximation method. Thus, the
approximation constants appeared in Barton et. al. method can also be
calculated by using the results of the exact method.
\end{abstract}

\section{Introduction}

In 1990, Barton et. al. had already published an approximation method to
calculate the energy shift in each quantum mechanical energy levels in a one
dimensional symmetrical linear harmonic oscillator. They found a formula of
the energy shifts either for the ground state or for the excited states
calculated in a finite interval $L$. The explanation about the use of their
approximation formula is for bound states in one dimensional symmetrical
linear harmonic oscillator with a potential $V\left( x\right) \,$which is
proportional to $x^2$.$^{\left[ 1\right] }$However promising their method,
the exact energy shifts in each energy levels have not been solved until now
yet. In this paper We use another method to solve the problem. Instead of
using the Barton et. al. approximation method, We apply a new method by
changing Schr\"{o}dinger differential equation to a confluent hypergeometric
differential equation. After finding the confluent hypergeometric
differential equation, We solve exactly the equation and then use a certain
program or software to calculate the final exact energy shifts.

This paper is organized as follows. In section 2, we will perform the exact
solution method of the energy shift in each quantum mechanical energy levels
in a one dimensional symmetrical linear harmonic oscillator and will also
compare our results with the results in ref.$[1]$. Section 3 is devoted for
discussions and conclusions.

\section{Exact solution method of the energy shift in each quantum
mechanical energy levels in a one dimensional symmetrical linear harmonic
oscillator}

A particle with mass $m$ moving in a one dimensional space from $-\infty $%
\thinspace to $\infty \,$with the influence of a potential $V\left( x\right)
\,$obeys the following Schr\"{o}dinger differential equation$^{[1]}$ 
\begin{equation}
\left( -\frac{\hbar ^2}{2m}\frac{d^2}{dx^2}\,+V\left( x\right) \right) \psi
_0\left( x\right) \,=E_0\psi _0\left( x\right) ,\,\,\,\,\,-\infty
\,<x<\infty \,.\,  \eqnum{2.1}
\end{equation}

On the other hand, if the motion of the particle is restricted by a finite
interval $-\frac L2$\thinspace $<x<\frac L2\,\,$in the same potential, the
wave function $\psi \left( x\right) $ and the energy $E$ can be obtained by
solving the following time independent Schr\"{o}dinger differential equation

\begin{equation}
\left( -\frac{\hbar ^2}{2m}\frac{d^2}{dx^2}\,+V\left( x\right) \right) \psi
\left( x\right) \,=E\psi \left( x\right) ,\,\,\,\,\,\,\,\,\,\,-\frac L2\,<x<%
\frac L2\,.  \eqnum{2.2}
\end{equation}
Even though the energy shift $\Delta E\equiv E-E_0$ is small for the large $%
L $, this case is not included in a classical perturbation theory.

Based on this simple physical phenomena, Barton et. al.$\,$derived generally
an approximation method for calculating the energy changes or the energy
shifts in each quantum mechanical energy levels including the systems in
linear harmonic oscillator and hydrogen atom. They got the following formulas

\begin{equation}
\Delta E\,=\,\frac{\hbar ^2}m\left[ \int\limits_0^{L/2}\frac{dx_1}{\psi
_0^2\left( x_1\right) }\right] ^{-1},\text{for ground state,}  \eqnum{2.3a}
\end{equation}
and

\begin{equation}
\Delta E\,\simeq \,\frac{\hbar ^2}m\left[ \int\limits_a^{L/2}\frac{dx_1}{%
\psi _0^2\left( x_1\right) }\right] ^{-1}\,(\,0<a<\frac L2,\text{\thinspace
and}\,x_1\sim \frac L2),  \eqnum{2.3b}
\end{equation}
for excited states. However, for the calculation of the energy shift in each
quantum mechanical energy levels in a one dimensional symmetrical linear
harmonic oscillator with a potential $V\left( x\right) =m\omega ^2x^2/2$,
they found

\begin{equation}
\frac{\Delta E^{\left( 0\right) }}{E_0}\,=\,\frac 2{\pi ^{1/2}}\left( \frac L%
l\right) \exp \left( -\frac{L^2}{4l^2}\right) \left( 1+O\left( \frac{l^2}{L^2%
}\right) \right) ,\,\,\,\,\text{for ground state}  \eqnum{2.4a}
\end{equation}
where $\psi _0\left( x\right) =\left( \frac 1{\pi ^{1/4}l^{1/2}}\right) \exp
\left( -\frac{x^2}{2l^2}\right) $,\thinspace $l=\left( \frac \hbar {m\omega }%
\right) ^{1/2}$, and\thinspace \thinspace the initial energy\thinspace is $%
E_0$\thinspace =$\frac 12\hbar \omega $. And for the excited states, they got

\begin{equation}
\frac{\Delta E^{\left( n\right) }}{E_0^{\left( n\right) }}\,=\,\frac 2{%
\left( 2n+1\right) \pi ^{1/2}n!2^n}\left( \frac Ll\right) ^{2n+1}\exp \left(
-\frac{L^2}{4l^2}\right) \left( 1+O\left( \frac{l^2}{L^2}\right) \right) , 
\eqnum{2.4b}
\end{equation}
where $\psi _0^{\left( n\right) }\left( x\right) =\left( \pi
^{1/2}l2^2n!\right) ^{-1/2}H_n\left( \frac xl\right) \exp \left( -\frac{x^2}{%
2l^2}\right) $and\thinspace $E_0^{\left( n\right) }$\thinspace =$\left( 
\frac 12+n\right) \hbar \omega $, $n=1,2,3,...\,$. For the large $x$, $%
H_n\left( \frac xl\right) \sim 2^n\left( \frac xl\right) ^n$.

Now, we are going to change the Schr\"{o}dinger differential equation to
become a confluent hypergeometric differential equation. We provide eigen
function $H\psi =E\psi \,$for a one dimensional symmetrical linear harmonic
oscillator with a potential $V\left( x\right) =m\omega ^2x^2/2\,$ into two
areas :$^{[2]}$

\begin{equation}
\left( -\frac{\hbar ^2}{2m}\frac{d^2}{dx^2}\,+\frac 12m\omega
_{+}^2x^2\right) \psi _{+}\left( x\right) \,=E\psi _{+}\left( x\right)
,\,\,\,\,\,\text{for}\,x>0,  \eqnum{2.5a}
\end{equation}
and

\begin{equation}
\left( -\frac{\hbar ^2}{2m}\frac{d^2}{dx^2}\,+\frac 12m\omega
_{-}^2x^2\right) \psi _{-}\left( x\right) \,=E\psi _{-}\left( x\right)
,\,\,\,\,\,\text{for}\,x<0.  \eqnum{2.5b}
\end{equation}

By substituting parameters $\frac{2mE}{\hbar ^2}=\lambda $,\thinspace and $%
\frac{m\omega }\hbar =\alpha \,$or\thinspace $\alpha =\frac 1{l^2}$, we find

\begin{equation}
\frac{d^2\psi _{\pm }}{dx^2}-\left[ \lambda -\alpha ^2x^2\right] \psi _{\pm
}=\,0,\,\text{for}\pm x>0.\,  \eqnum{2.6}
\end{equation}
We then introduce parameter $\lambda =\alpha \left( 2n^{\prime }+1\right) $
and a free variable $z=x\sqrt{\alpha }\,$ to eq.(2.6), we get

\begin{equation}
\frac{d^2\psi _{n^{\prime }}\left( z\right) }{dz^2}-\left[ 2n^{\prime
}+1-z^2\right] \psi _{n^{\prime }}\left( z\right) =\,0,\text{\thinspace for}%
\pm z>0,\,  \eqnum{2.7}
\end{equation}
and

\begin{equation}
E_{n^{\prime }}=\hbar \omega \left( \frac 12+n^{\prime }\right) . 
\eqnum{2.8}
\end{equation}
It means that

\begin{equation}
\frac{\omega _{-}}{\omega _{+}}=\frac \omega \omega =\frac{2n^{\prime }+1}{%
2n^{\prime }+1}=r=1,  \eqnum{2.9}
\end{equation}
which states that our solution is for a one dimensional symmetrical linear
harmonic oscillator.

If $\psi _{n^{\prime }}\left( z\right) \,$is the solution of eq.(2.7), then $%
\psi _{n^{\prime }}\left( -z\right) \,$is also the solution. On the other
hand, if the Wronskian $W\left( \psi _{n^{\prime }}\left( x\sqrt{\alpha }%
\right) ,\psi _{n^{\prime }}\left( -x\sqrt{\alpha }\right) \right) \,$is not
the same as zero, then both solutions can be used as a basic solution for
implementing boundary conditions whether in $x\rightarrow \pm \frac L2$
\thinspace or in $x=0$. To solve eq.(2.7), we firstly substitute the
following function

\begin{equation}
\psi _{n^{\prime }}\left( z\right) =\exp \left( -\frac 12z^2\right)
H_{n^{\prime }}\left( z\right) ,  \eqnum{2.10}
\end{equation}
we then get

\begin{equation}
\left[ \frac{d^2}{dz^2}-2z\left( \frac d{dz}\right) -2n^{\prime }\right]
H_{n^{\prime }}\left( z\right) =0,  \eqnum{2.11}
\end{equation}
where $H_{n^{\prime }}\left( z\right) \,$is a Hermite function. By
introducing a new variable $t=z^2$, we find a confluent hypergeometric {\bf %
(CH) }differential equation

\begin{equation}
\left[ t\frac{d^2}{dz^2}-\left( \frac 12-t\right) \left( \frac d{dt}\right) -%
\frac 12n^{\prime }\right] H_{n^{\prime }}\left( t\right) =0,  \eqnum{2.12}
\end{equation}
which has a general solution in a linear combination of the following two 
{\bf CH }functions$^{[3]}$

\begin{equation}
A\,\,_1F_1\left( -\frac{n^{\prime }}2;\frac 12;t=z^2\right) +B\left( \sqrt{t}%
\right) \,\,_1F_1\left( \frac{1-n^{\prime }}2;\frac 32;t=z^2\right) . 
\eqnum{2.13}
\end{equation}
In eq.(2.10), the Hermite function $H_{n^{\prime }}\left( z\right) \,$is
just a special function of the following equation

\begin{equation}
H_{n^{\prime }}\left( z\right) =\frac{2^{n^{\prime }}\Gamma \left( \frac 12%
\right) }{\Gamma \left( \frac 12\left( 1-n^{\prime }\right) \right) }%
\,_1F_1\left( -\frac{n^{\prime }}2;\frac 12;z^2\right) +\,\frac{2^{n^{\prime
}}\Gamma \left( -\frac 12\right) }{\Gamma \left( -\frac{n^{\prime }}2\right) 
}\,\left( z\right) _1F_1\left( \frac{1-n^{\prime }}2;\frac 32;z^2\right) . 
\eqnum{2.14}
\end{equation}
Here, we have chosen constants $A$ and $B$ as follows$^{[4]}$

\begin{equation}
A=\frac{2^{n^{\prime }}\Gamma \left( \frac 12\right) }{\Gamma \left( \frac 12%
\left( 1-n^{\prime }\right) \right) },  \eqnum{2.15a}
\end{equation}
and

\begin{equation}
B=\frac{2^{n^{\prime }}\Gamma \left( -\frac 12\right) }{\Gamma \left( -\frac{%
n^{\prime }}2\right) }.  \eqnum{2.15b}
\end{equation}
The boundary conditions in $z$ are

\begin{equation}
H_{n^{\prime }}\left( 0\right) =\frac{2^{n^{\prime }}\Gamma \left( \frac 12%
\right) }{\Gamma \left( \frac 12\left( 1-n^{\prime }\right) \right) }, 
\eqnum{2.16a}
\end{equation}
and

\begin{equation}
\frac{dH_{n^{\prime }}\left( 0\right) }{dz}=\frac{2^{n^{\prime }}\Gamma
\left( -\frac 12\right) }{\Gamma \left( -\frac{n^{\prime }}2\right) }=\frac{%
-2^{n^{\prime }+1}\sqrt{\pi }}{\Gamma \left( -\frac{n^{\prime }}2\right) }. 
\eqnum{2.16b}
\end{equation}
We can also show that$^{[3]}$

\begin{equation}
W\left( \psi _{n^{\prime }}\left( x\sqrt{\alpha }\right) ,\psi _{n^{\prime
}}\left( -x\sqrt{\alpha }\right) \right) =\frac{2^{n^{\prime }}\sqrt{\pi }}{%
\Gamma \left( -n^{\prime }\right) }\exp \left( z^2\right) .  \eqnum{2.17}
\end{equation}
Hence, if $n^{\prime }$\thinspace is not exactly the same as $n^{\prime
}=1,2,3,...\,$,\thinspace then both of the solutions $H_{n^{\prime }}\left(
\pm z\right) \,\,$have a linear independent property. If this is work, then
both of the solutions can be chosen as a basic solution of eq.(2.7), we get
the general solution as

\begin{equation}
\psi _{n^{\prime }}\left( z\right) =\left[ AH_{n^{\prime }}\left( z\right)
+BH_{n^{\prime }}\left( -z\right) \right] \exp \left( -\frac 12z^2\right) . 
\eqnum{2.18}
\end{equation}
Here, the boundary condition for the eigen function in $\pm z=\pm \frac L2%
\sqrt{\alpha }$ is

\begin{equation}
\psi _{n^{\prime }}\left( \pm z\right) =\psi _{n^{\prime }}\left( \pm \frac L%
2\sqrt{\alpha }\right) =0.  \eqnum{2.19}
\end{equation}
By substituting the boundary condition in eq.(2.19) into eq.(2.18), we find

\begin{equation}
\psi _{n^{\prime }}\left( \pm \frac L2\sqrt{\alpha }\right) =\left[
AH_{n^{\prime }}\left( +\frac L2\sqrt{\alpha }\right) +BH_{n^{\prime
}}\left( -\frac L2\sqrt{\alpha }\right) \right] \exp \left( -\frac{L^2}8%
\alpha \right) =0.  \eqnum{2.20}
\end{equation}
Eq.(2.20) can be divided into two equations with two conditions as follows

$\left( 1\right) $. For the condition with even parity:

\[
\psi _{n^{\prime }}\left( \frac L2\sqrt{\alpha }\right) =\left[
AH_{n^{\prime }}\left( +\frac L2\sqrt{\alpha }\right) +BH_{n^{\prime
}}\left( -\frac L2\sqrt{\alpha }\right) \right] \exp \left( -\frac{L^2}8%
\alpha \right) =0, 
\]

or

\begin{equation}
2\left[ \frac{2^{n^{\prime }}\Gamma \left( \frac 12\right) }{\Gamma \left( 
\frac 12\left( 1-n^{\prime }\right) \right) }\,_1F_1\left( -\frac{n^{\prime }%
}2;\frac 12;\frac{L^2}4\alpha \right) \right] =0.  \eqnum{2.21a}
\end{equation}

$\left( 2\right) $. For the condition with odd parity:

\[
\psi _{n^{\prime }}\left( \frac L2\sqrt{\alpha }\right) =\left[
AH_{n^{\prime }}\left( +\frac L2\sqrt{\alpha }\right) -BH_{n^{\prime
}}\left( -\frac L2\sqrt{\alpha }\right) \right] \exp \left( -\frac{L^2}8%
\alpha \right) =0, 
\]

or

\begin{equation}
2\left[ \frac{2^{n^{\prime }}\Gamma \left( -\frac 12\right) }{\Gamma \left( -%
\frac{n^{\prime }}2\right) }\left( \frac L2\sqrt{\alpha }\right)
\,_1F_1\left( \frac{1-n^{\prime }}2;\frac 32;\frac{L^2}4\alpha \right)
\right] =0.  \eqnum{2.21b}
\end{equation}

In both eq.(2.21a) and (2.21b), hermite functions $H_{n^{\prime }}\left( \pm 
\frac L2\sqrt{\alpha }\right) $are$^{[5]}$

\begin{eqnarray}
H_{n^{\prime }}\left( +\frac L2\sqrt{\alpha }\right) &=&\left[ \frac{%
2^{n^{\prime }}\Gamma \left( \frac 12\right) }{\Gamma \left( \frac 12\left(
1-n^{\prime }\right) \right) }\,_1F_1\left( -\frac{n^{\prime }}2;\frac 12;%
\frac{L^2}4\alpha \right) \right]  \nonumber \\
&&+\left[ \frac{2^{n^{\prime }}\Gamma \left( -\frac 12\right) }{\Gamma
\left( -\frac{n^{\prime }}2\right) }\left( \frac L2\sqrt{\alpha }\right)
\,_1F_1\left( \frac{1-n^{\prime }}2;\frac 32;\frac{L^2}4\alpha \right)
\right] ,  \eqnum{2.22a}
\end{eqnarray}
and

\begin{eqnarray}
H_{n^{\prime }}\left( -\frac L2\sqrt{\alpha }\right) &=&\left[ \frac{%
2^{n^{\prime }}\Gamma \left( \frac 12\right) }{\Gamma \left( \frac 12\left(
1-n^{\prime }\right) \right) }\,_1F_1\left( -\frac{n^{\prime }}2;\frac 12;%
\frac{L^2}4\alpha \right) \right]  \nonumber \\
&&-\left[ \frac{2^{n^{\prime }}\Gamma \left( -\frac 12\right) }{\Gamma
\left( -\frac{n^{\prime }}2\right) }\left( \frac L2\sqrt{\alpha }\right)
\,_1F_1\left( \frac{1-n^{\prime }}2;\frac 32;\frac{L^2}4\alpha \right)
\right] ,  \eqnum{2.22b}
\end{eqnarray}
where

\begin{eqnarray}
_1F_1\left( -\frac{n^{\prime }}2;\frac 12;\frac{L^2}4\alpha \right) &=&\frac{%
\Gamma \left( \frac 12\right) }{\Gamma \left( -\frac{n^{\prime }}2\right) }%
%TCIMACRO{\dsum }
%BeginExpansion
\mathop{\displaystyle \sum }
%EndExpansion
\limits_{s=1}^\infty \frac{\Gamma \left( -\frac{n^{\prime }}2+s\right)
\left( \frac{L^2}4\alpha \right) ^s}{\Gamma \left( \frac 12+s\right) s!} 
\nonumber \\
&=&1-\frac{n^{\prime }L^2\alpha }4+\frac{\left( n^{\prime 2}-2n^{\prime
}\right) L^4\alpha ^2}{96}+...\,\,\,,  \eqnum{2.23a}
\end{eqnarray}
and

\begin{eqnarray}
_1F_1\left( \frac{1-n^{\prime }}2;\frac 32;\frac{L^2}4\alpha \right) &=&%
\frac{\Gamma \left( \frac 32\right) }{\Gamma \left( \frac{\left( 1-n^{\prime
}\right) }2\right) }%
%TCIMACRO{\dsum }
%BeginExpansion
\mathop{\displaystyle \sum }
%EndExpansion
\limits_{s=1}^\infty \frac{\Gamma \left( \frac{1-n^{\prime }}2+s\right)
\left( \frac{L^2}4\alpha \right) ^s}{\Gamma \left( \frac 32+s\right) s!} 
\nonumber \\
&=&1+\frac{\left( 1-n^{\prime }\right) L^2\alpha }{12}+\frac{\left(
1-n^{\prime }\right) \left( 3-n^{\prime }\right) L^4\alpha ^2}{480}%
+...\,\,\,.  \eqnum{2.23b}
\end{eqnarray}
From eq.(2.21b), we find a solution for the odd parity,

\begin{equation}
n^{\prime }=1+2m^{\prime }.  \eqnum{2.24a}
\end{equation}
On the other hand, the even parity is found by calculating eq.(2.21a), we
then get 
\begin{equation}
n^{\prime }=2m^{\prime }.  \eqnum{2.24b}
\end{equation}
Here (in eq.(2.24a) and (2.24b)), $n^{\prime }\,$and $m^{\prime }\,$are
\thinspace positive real values due to the boundary condition $z=\pm \frac L2%
\sqrt{\alpha }\,$where $\psi _{n^{\prime }}\left( \pm z\right) =0\,$vanishes
in $\,z=\pm \frac L2\sqrt{\alpha }$.$\,\,n^{\prime }\,$in eq.(2.21a) and
(2.21b) can exactly be calculated by using mathcad plus 6.0 or a program
programmed by using Turbo Pascal 7.0. The results of the calculation are
provided in table 1 and 2.

The requirement in eq.(2.24b) produces the bound states quantum mechanical
energy with even parity,

\begin{equation}
E_{n^{\prime }}=\left( n^{\prime }+\frac 12\right) \hbar \omega
,\,\,\,\,\,n^{\prime }=2m^{\prime }\,,  \eqnum{2.25a}
\end{equation}
and the requirement in eq.(2.24a) produces the bound states quantum
mechanical energy with odd parity as follows

\begin{equation}
E_{n^{\prime }}=\left( n^{\prime }+\frac 12\right) \hbar \omega
,\,\,\,\,\,n^{\prime }=1+2m^{\prime }.\,  \eqnum{2.25b}
\end{equation}

On the other hand, for the calculation of the quantum mechanical energy in
the boundary $z=\pm \infty $,$\,$and the boundary condition $\psi _n\left(
\pm z\right) =$finite, we can use an asymtotic formula of $H_n\left( \pm
z\right) \,\,$according to ref.$[3]$ as follows

\begin{equation}
H_n\left( \pm z\right) \simeq \left( \pm 2z\right) ^n,\,\text{for}%
\,\,\,z\rightarrow \pm \infty \,\,\,\text{or\thinspace }\,\left| z\right|
\rightarrow \infty \,\,\,\text{if\thinspace \thinspace \thinspace }\left|
\arg \left( z\right) \right| <\frac{3\pi }4,  \eqnum{2.26a}
\end{equation}
and

\begin{equation}
H_n\left( \pm z\right) \simeq \left( \sqrt{\pi }\right) \exp \left(
z^2\right) \left( \left| z\right| \right) ^{-n-1},\,\text{for }z\rightarrow
\pm \infty .  \eqnum{2.26b}
\end{equation}
By substituting the conditions in eq.(2.26a) and (2.26b) into eq.(2.18), we
find the bound states quantum mechanical energy with even parity,

\begin{equation}
E_n=\left( n+\frac 12\right) \hbar \omega
,\,\,\,\,\,n=2m=0,2,4,6...\,\,\,\,\,,  \eqnum{2.27a}
\end{equation}
and for the odd parity, we get

\begin{equation}
E_n=\left( n+\frac 12\right) \hbar \omega ,\,\,\,\,\,n=1+2m=1,3,5,...\,\,.\,
\eqnum{2.27b}
\end{equation}
Hence, for a one dimensional symmetrical linear harmonic oscillator in the
boundary $z=\pm \infty \,$, we get its wave function as

\begin{equation}
\psi _n\left( x\right) =\left[ A_nH_n\left( x\sqrt{\alpha }\right) \right]
\exp \left( -\frac 12\alpha x^2\right) ,  \eqnum{2.28}
\end{equation}
where the constant parameter $A_n\,$can be found by a normalization.

Now, the energy shift in each quantum mechanical energy levels in the one
dimensional symmetrical linear harmonic oscillator can exactly be calculated
by formulating a simple formula related to eq.(2.25) and eq.(2.27) as follows

\begin{eqnarray}
\Delta E^{(n)} &=&E_{n^{\prime }}-E_n  \nonumber \\
&=&\left( n^{\prime }-n\right) \hbar \omega ,  \eqnum{2.29}
\end{eqnarray}
where $n=0,1,2,3,4,5,...$(positive integer)\thinspace and the values of
parameter $n^{\prime }\,$depend on $\frac L2\sqrt{\alpha }$, for example :
for $\frac L2\sqrt{\alpha }=6$,

\[
n^{\prime }=1.55\text{x}10^{-15},\left( 1+1.082\text{x}10^{-13}\right)
,\left( 2+3.671\text{x}10^{-12}\right) ,\left( 3+0.805\text{x}%
10^{-10}\right) ,\text{etc.} 
\]
Some results of the energy shift are attached in table 1 and 2.

\thinspace \thinspace \thinspace \thinspace \thinspace \thinspace \thinspace
\thinspace 
\begin{tabular}{||l||l||l||}
\hline\hline
& \thinspace \thinspace \thinspace \thinspace \thinspace Barton et. al.
results (1990) & \thinspace \thinspace \thinspace \thinspace \thinspace
\thinspace \thinspace \thinspace \thinspace \thinspace \thinspace \thinspace
Our results \\ 
$k=\frac L2\sqrt{\alpha }$ & $\,\,\,\,\left( \text{see}:\text{eq.}%
(2.4a),\,\,\Delta E^{(0)}\right) $ & $\,\,\,\,\,\,\left( \Delta
E^{(0)}=E_{0^{\prime }}-E_0\right) $ \\ 
&  &  \\ \hline\hline
1 & \thinspace \thinspace \thinspace \thinspace 0.415$\left( 1+O\left( \frac 
14\right) \right) \,\hbar \omega $ & \thinspace \thinspace \thinspace
\thinspace \thinspace \thinspace \thinspace 0.798\thinspace $\hbar \omega $
\\ \hline\hline
3 & \thinspace \thinspace \thinspace \thinspace 4.177x10$^{-4}\left(
1+O\left( \frac 1{36}\right) \right) \,\hbar \omega $ & \thinspace
\thinspace \thinspace \thinspace \thinspace \thinspace \thinspace 3.911x10$%
^{-4}$ $\,\hbar \omega $ \\ \hline\hline
4 & \thinspace \thinspace \thinspace \thinspace 5.079x10$^{-7}\left(
1+O\left( \frac 1{64}\right) \right) \,\hbar \omega $ & \thinspace
\thinspace \thinspace \thinspace \thinspace \thinspace \thinspace 4.908x10$%
^{-7}$ $\,\hbar \omega $ \\ \hline\hline
5 & \thinspace \thinspace \thinspace \thinspace 7.835x10$^{-11}\left(
1+O\left( \frac 1{100}\right) \right) \,\hbar \omega $ & \thinspace
\thinspace \thinspace \thinspace \thinspace \thinspace \thinspace 7.671x10$%
^{-11}\,\hbar \omega $ \\ \hline\hline
6 & \thinspace \thinspace \thinspace \thinspace 1.570x10$^{-15}\left(
1+O\left( \frac 1{144}\right) \right) \,\hbar \omega $ & \thinspace
\thinspace \thinspace \thinspace \thinspace \thinspace \thinspace 1.550x10$%
^{-15}\,\hbar \omega $ \\ \hline\hline
7 & \thinspace \thinspace \thinspace \thinspace 4.141x10$^{-21}\left(
1+O\left( \frac 1{196}\right) \right) \,\hbar \omega $ & \thinspace
\thinspace \thinspace \thinspace \thinspace \thinspace \thinspace 4.098x10$%
^{-21\,}\hbar \omega $ \\ \hline\hline
10 & \thinspace \thinspace \thinspace \thinspace 4.197x10$^{-43}\left(
1+O\left( \frac 1{400}\right) \right) \,\hbar \omega $ & \thinspace
\thinspace \thinspace \thinspace \thinspace 36.769x10$^{-43}\hbar \omega $
\\ \hline\hline
\end{tabular}

Table 1. The\thinspace energy shift in the ground state of the one
dimensional symmetrical linear harmonic oscillator.

\thinspace \thinspace \thinspace \thinspace \thinspace \thinspace \thinspace
\thinspace 
\begin{tabular}{||l||l||l||}
\hline\hline
& \thinspace \thinspace \thinspace \thinspace Barton et. al. results (1990)
& \thinspace \thinspace \thinspace \thinspace Our results \\ 
$k=\frac L2\sqrt{\alpha }$ & $\,\,\,\,\left( \text{see}:\text{eq.}%
(2.4b),\,\,\Delta E^{(n)}\right) $ & $\,\,\,\,\left( \Delta
E^{(n)}=E_{n^{\prime }}-E_n\right) $ \\ 
&  &  \\ \hline\hline
3 & $\,\,\,\,\Delta E^{(1)}=\,$7.52x10$^{-3}A\,\hbar \omega $ & $\Delta
E^{(1)}=\,$6.081x10$^{-3}\,\hbar \omega $ \\ 
& $\,\,\,\,\Delta E^{(2)}=\,$6.767x10$^{-2}A\,\hbar \omega $ & $\Delta
E^{(2)}=\,$\thinspace 4.119x10$^{-2}\hbar \omega $ \\ 
& \thinspace \thinspace \thinspace \thinspace where $A=\left( 1+O\left( 
\frac 1{36}\right) \right) $ & we get $A=\,$0.606 \\ 
&  &  \\ \hline\hline
5 & $\,\,\,\,\Delta E^{(1)}=\,$3.918x10$^{-9}B\,\hbar \omega $ & $\Delta
E^{(1)}=\,$3.672x10$^{-9}\,\hbar \omega $ \\ 
& $\,\,\,\,\Delta E^{(2)}=\,$9.794x10$^{-8}B\,\hbar \omega $ & $\Delta
E^{(2)}=\,$8.402x10$^{-8}\,\hbar \omega $ \\ 
& $\,\,\,\,\Delta E^{(3)}=\,$1.632x10$^{-6}B\,\hbar \omega $ & $\Delta
E^{(3)}=\,$1.221x10$^{-6}\,\hbar \omega $ \\ 
& \thinspace \thinspace \thinspace \thinspace where $B=\left( 1+O\left( 
\frac 1{100}\right) \right) $ & we get $B=\,$0.848 \\ 
&  &  \\ \hline\hline
6 & $\,\,\,\,\Delta E^{(1)}=\,$1.13x10$^{-13}C\,\hbar \omega $ & $\Delta
E^{(1)}=\,$1.08x10$^{-13}\,\hbar \omega $ \\ 
& $\,\,\,\,\Delta E^{(2)}=\,$4.07x10$^{-12}C\,\hbar \omega $ & $\Delta
E^{(2)}=\,$3.67x10$^{-12}\,\hbar \omega $ \\ 
& $\,\,\,\,\Delta E^{(3)}=\,$0.98x10$^{-10}C\,\hbar \omega $ & $\Delta
E^{(3)}=\,$0.80x10$^{-10}\,\hbar \omega $ \\ 
& \thinspace \thinspace \thinspace \thinspace where $C=\left( 1+O\left( 
\frac 1{144}\right) \right) $ & we get $C=\,$0.892 \\ \hline\hline
\end{tabular}

Table 2. The\thinspace energy shift in the excited states of the one
dimensional symmetrical linear harmonic oscillator.

\section{Discussions and Conclusions}

We have presented an exact solution method of the energy shift in each
quantum mechanical energy levels in a one dimensional symmetrical linear
harmonic oscillator by using the solution of the confluent hypergeometric
differential equation. We can conclude that the relationship of the energy
shifts is as follows :

\[
\Delta E^{(0)}<\Delta E^{(1)}<\Delta E^{(2)}<\Delta E^{(3)}<...<\Delta
E^{(n-1)}<\Delta E^{(n)}. 
\]

According to the comparison between our results and Barton et. al. results
shown in table 1 and 2, we get that their results are almost the same as our
results. However, our results are the exact results. On the other hand, our
results will be more accurate than their method if in our calculation, we
add the series of gamma function in eq.(2.23a) and (2.23b).The longer gamma
function series, the more accurate our results. Based on the comparison, we
also get the approximation constants (obliquity factors) appeared in Barton
et. al. method.

In conclusion, our method is work without a detail calculation of the wave
function in each energy levels.

${\bf Acknowledgements}$

We would like to thank G.Barton for useful discussions. We also would like
to thank K.Esomar for his encouragement. The work of H.E. was supported by
the PDM No.170/P2IPT/LITMUD/V/1997, Ditbinlitabnas-DIKTI, Republic of
Indonesia.

\[
\]

\end{document}